\documentclass[12pt,a4paper]{article}
\usepackage{amsmath}
\usepackage{graphicx}
\usepackage{times}
\begin{document}
\begin{titlepage}
\title{Analytization of elastic scattering amplitude}
\author{ S.M. Troshin$^*$, N.E. Tyurin\\[1ex]
\small  \it SRC IHEP of NRC ``Kurchatov Institute''\\
\small  \it Protvino, 142281, Russian Federation\\[1ex]
{\small\it  *) Sergey.Troshin@ihep.ru}}
\normalsize
\date{}
\maketitle

\begin{abstract}
Scenario for  restoration of the real part of the elastic scattering amplitude has been proposed for  the unitarity saturation case.
Dependence of the real part of the elastic scattering amplitude on the transferred momentum $-t$ at the asymptotical energies has been restored from the corresponding imaginary part on the basis of derivative analyticity relations (analytization procedure).\\[1ex]

\end{abstract}
\end{titlepage}
\section{Introduction}
\label{intro}
The recent experimental results of the TOTEM Collaboration  on the measurements of the differential cross-section of elastic scattering $pp$--scattering at small $-t$
  at the energy in the center of mass system $\sqrt{s}=8$ TeV \cite {totem} have demonstrated strong deviation  of the nuclear amplitude from a simple exponential dependence. Those measurements have  high precision and significance of the above effect is at the level of seven standards.  We discuss here the role of the real part for the explanation of this deviation and confirm the possibility of emerging $-t$ dependence of the ratio of  real to imaginary parts $\rho(s,t)$ of the elastic scattering amplitude at small $-t$ values in case of the unitarity limit saturation  at  the asymptotical values of the energy $s\to\infty$. 
 
 The problem  has a long story. A knowledge of the real part of the elastic  amplitude is particularly important due to its sensitivity to the possible dispersion relation violation resulting from violation of the polynomial boundedness \cite{martin} and to the new physics  related, e.g. to the non-local interactions resulting from presence of the fundamental length and/or existence of an extra internal compact dimensions \cite{khuri,bour}. However, this knowledge is rather limited nowadays: the experimental data provide  information for the forward scattering in the Coulomb-Nuclear Interference (CNI) region only \cite{totcni}, while it could play an important role in a wider region of the transferred momentum $t$ variation. It should be noted that the real part of the scattering amplitude has a peripheral dependence on impact parameter in the both cases of black-disk or unitarity limit saturation (cf. \cite{trosh,anis2}). It means that if the asymptotics corresponds to one of the above limits the function $\rho(s,t)$ has a nontrivial $t$ dependence at $s\to\infty$.
 
 The black disk limit for the scattering amplitude in the impact parameter representation (in case the pure imaginary amplitude) is $1/2$ and unitarity limit is unity.
The ``black disk model'' implies that the amplitude in the impact parameter representation is equal to $1/2$ till some value of the impact parameter $b=R$ and it decreases at $b>R$. Similar form of this dependence is valid for the case of unitarity limit saturation which is twice as much as the black disk limit. It should be noted, that the asymptotic equipartition $\sigma_{el}/\sigma_{tot}=\sigma_{inel}/\sigma_{tot}=1/2$ does not define the ``black disk model'' since the same relation is being valid in various cases.  The forward scattering observables are the definite integrals over impact parameter and it is quite evident,  that an integration over impact parameter does not allow an unequivocal reconstruction of an initial integrand. The above equipartition  at 
$s \to \infty$ has been obtained in  \cite{white}  for a gaussian exponential dependence of the profile function saturating unitarity limit at $b=0$, and this function has nothing to do with the black disc model.  

 The dominance of the imaginary part of the scattering amplitude is almost commonly accepted in the region of diffraction cone, but the real part should not be neglected in the whole region of the transferred momentum variation since it is related through the dispersion relations to the imaginary part of the scattering amplitude and therefore it is not an arbitrary function. We assume the validity of dispersion relations and, more generally, validity of the anylyticity, unitarity and polynomial boundedness and then restore the real part of the scattering amplitude  using the derivative analyticity relations \cite{anis2,grib,bronz,bronz1}. Those local relations have been obtained  in the limit of $s\to\infty$. 
 
 The following scenario for construction of the real and imaginary parts of the scattering amplitude  can be adopted. Namely,  the procedure is used: we calculate  imaginary part of the scattering amplitude using unitarization scheme based on rational representation  where an input is taken to be a pure imaginary and real part of the input is neglected at this first stage. At the second stage the real part of the amplitude is restored from the imaginary one on the basis of the derivative analyticity relations. This stage is proposed to be  called as ``analytization'' by analogy with the term unitarization. The final amplitude includes both imaginary and real parts and is consistent with unitarity; the limit $\mbox{Im} f\to 1$ implies that   
 $\mbox{Re} f\to 0$
 at $s\to\infty$ and  fixed impact parameter. Normalization  of the amplitude $f$ is determined by the unitarity relation
 \begin{equation}\label{unrel}
 \mbox{Im} f(s,b)=h_{el}(s,b)+h_{inel}(s,b),
 \end{equation}
 where $h_{el,inel}$ are the elastic and inelastic overlap functions, respectively and $h_{el}\equiv |f(s,b)|^2$.  As it will be shown below, the limit $\mbox{Re} f\to 0$ at $s\to\infty$ at any fixed value of $b$ results from its
  proportionality 
 to the inelastic overlap function $h_{inel}$ which tends to zero at $s\to \infty$ at any fixed value of the impact parameter.
 
\section{Real part of the elastic scattering amplitude and unitarity saturation}
\label{sec:1}
Till the measurements at the LHC energies, the elastic scattering  data have been consistent with the BEL picture, i.e. when the protons' interaction region becomes Blacker (increase of absorption) at the center, relatively Edgier and Larger.  The analysis of the data on elastic scattering obtained by the TOTEM  at the energy $\sqrt{s}=7$ TeV has revealed definite hints on the presence  of the  new mode in strong interaction dynamics  \cite{ph1,phl,degr,intja,anis}  called as
antishadowing, reflective or resonant mode. It constitutes   a gradual transition to the so called REL picture, i.e. when the elastic interaction  starts to  be Reflective (i.e. the corresponding scattering matrix element becomes negative) at the center and simultaneously appears to be relatively Edgier, Larger and blacker at its  periphery. Asymptotic picture can then be described as a black ring, its possible appearance has been mentioned in \cite{phl}. 

There are  phenomenological models which can successfully describe the existing experimental data (cf. e.g.  \cite{fagundes} and the references therein).

The appearance of the reflective mode is registered under reconstruction of the elastic amplitude, elastic and inelastic
overlap functions in the impact parameter representation  \cite{alkin}. It is turned out that the scattering amplitude in the impact parameter representation is monotonically reaching the black-disk limit $1/2$ from below. It allows one to assume a monotonic increase with  energy in case  of saturation of the unitarity limit by the elastic scattering amplitude in the impact parameter representation. 
We follow here  an opportunity related to the unitarity limit saturation. It was  supported by the analysis of the experimental data on elastic scattering in the impact parameter representation \cite{alkin}. The recent
luminosity-independent measurements  at $\sqrt{s}= 8$ TeV \cite{8tev} have confirmed an increase of the ratio $\sigma_{el}(s)/\sigma_{tot}(s)$, which is another, but indirect, indication on the reflective scattering mode presence.  

Discussions of the possible interpretations 
of this mode based on the quark-gluon structure of the hadrons can be found in \cite{refl,drem,granad}.

 To restore the real part of the scattering amplitude we use derivative analyticity relation in the impact parameter representation.  The most simple and straightforward derivation of this relation in the impact parameter representation has been given recently in
 \cite{anis2} under assumption of the dominating imaginary part of the scattering amplitude:
\begin{equation}\label{darb}
\mbox{Re} f(s,b)\simeq \frac{\pi}{2}\frac{\partial \mbox{Im} f(s,b)}{\partial \ln s}.
\end{equation}
The amplitudes of $pp$ and $\bar{p} p$ are taken to be equal. This implies an absence of the crossing-odd contributions like the ones generated by the odderon at $s\to\infty$ \cite{trosh}.
Such relation (integrated over $b$) has been used in \cite{ph1} for analysis of the energy dependence of the parameter $\rho (s)
\equiv \rho(s,t)|_{-t=0}$, where $\rho(s,t)\equiv \mbox{Re}F(s,t)/\mbox{Im}F(s,t)$.
The obtained energy dependence has been found to obey the Khuri-Kinoshita theorem \cite{kinosh}, i.e. it decreases like $\pi /\ln s$ 
if $\sigma_{tot}(s)\sim \ln^2 s$ at  $s\to\infty$ . Here we extend the above consideration to the case of $-t\neq 0$.

To construct imaginary part of the scattering amplitude in the impact parameter representation we use the rational form of unitarization, its origin can be traced back to the papers of Heitler \cite{heitler}, and assume first that the input function $U$ is pure imaginary, i.e. use the replacement $U\to iU$. It gives a pure real elastic scattering matrix element $S(s,b)$ and pure real scattering amplitude with the same replacement $f\to if$. The corresponding relations have the forms
\begin{equation}\label{ssb}
S(s,b)=1-2f(s,b),
\end{equation}
\begin{equation}\label{fsb}
f(s,b)=\frac{U(s,b)}{1+U(s,b)}
\end{equation}
and
\begin{equation}\label{ssbr}
S(s,b)=\frac{1-U(s,b)}{1+U(s,b)}.
\end{equation}
We  introduce the function $r(s)$ which is determined as a solution of the equation $U(s,b)=1$, i.e $U(s,b=r(s))=1$. Under this, we suppose that the function $U(s,b)$ monotonically increases with the energy at any fixed impact parameter $b$. This increase is supposed to be a power-like one. Such a dependence  leads to rising behavior of the total cross-section like $\ln^2 s$ at $s\to\infty$ and unitarity saturation, i.e $f(s,b)\to 1$ at any fixed $b$ when $s\to \infty $. The observed increase of $\sigma_{tot}(s)$ can be considered as an argument in favor of the supposed dependence of the function $U(s,b)$. When the function $U$ becomes greater than unity, the reflective scattering mode starts to appear \cite{mpla16}. The essential feature of this  mode is a peripheral distribution  of the inelastic production probability over the impact parameter with maximum reached at $b=r(s)$. The peripheral form becomes more and more prominently peaked with an energy increase.

In our model, we use factorized form for the $s$ and $b$ dependencies of the function $U(s,b)$ with an explicit power-like energy increasing dependence of $U(s,b)$ and linear exponential decrease of this function with the impact parameter $b$
 (i.e. with an energy-independent slope).
The corresponding asymptotical dependencies  $r(s) \sim \ln s$ and $\sigma_{inel}(s) \sim \ln s$, while  $\sigma_{el}(s) \sim \ln^2 s$ (cf. e.g. \cite{usmpla}).
 
 Now, we can calculate $\mbox{Re} f(s,b)$ according to Eqs. (\ref{darb}) and (\ref{fsb}):
 \begin{equation}\label{rfsb}
 \mbox{Re} f(s,b)=\frac{\pi}{2}\frac{\partial \ln U(s,b)}{\partial \ln s}h_{inel}(s,b),
 \end{equation}
 where
 \begin{equation}\label{hin}
 h_{inel}(s,b)=\frac{U(s,b)}{[1+U(s,b)]^2}
 \end{equation}
 is the inelastic overlap function. Probability distribution of the inelastic processes over impact parameter is $ P_{inel}(s,b)=4h_{inel}(s,b)$.
  The real part of the scattering amplitude $\mbox{Re} F(s,t)$ can be calculated as the Fourier-Bessel transform of $f(s,b)$:
 \begin{equation}\label{rfst}
 \mbox{Re} F(s,t)=\frac{s}{\pi^2}\int_0^{\infty}bdb\mbox{Re} f(s,b)J_0(b\sqrt{-t}).
 \end{equation}
We have already mentioned that at asymptotical energies inelastic overlap function $h_{inel}(s,b)$ has a very prominent maximum at $b=r(s)$ and tends to zero at $s\to\infty$ and $b=0$. Such peripheral form with a peak at $b=r(s)$ allows one to obtain
an approximate relation
\begin{equation}\label{rfst1}
 \mbox{Re} F(s,t)\simeq\frac{s}{16\pi^2}\left.\frac{\partial \ln U(s,b)}{\partial \ln s}\right |_{b=r(s)}\sigma_{inel}(s)J_0(r(s)\sqrt{-t}).
 \end{equation}
Since for the function $U(s,b)$ the factorized dependence on the variables $s$ and $b$ with power-like energy dependence was taken, we have 
\begin{equation}\label{der}
\left.\frac{\partial \ln U(s,b)}{\partial \ln s}\right |_{b=r(s)} =\mbox{const.}
\end{equation}
It should be noted, that due to Eq. (\ref{der}), the following relations take place in the case of the unitarity limit saturation:
\begin{equation}\label{derb}
{\mbox{Re} f(s,b)} \sim h_{inel}(s,b)
\end{equation}
and
\begin{equation}\label{dert}
\mbox{Re} F(s,t) \sim H_{inel}(s,t),
\end{equation}
where
\begin{equation}\label{hinst}
 H_{inel}(s,t)=\frac{s}{\pi^2}\int_0^{\infty}bdbh_{inel}(s,b)J_0(b\sqrt{-t}).
 \end{equation}
 
There are no such relations for the case of the black disk limit saturation, when the inelastic overlap function is not peripheral at $s\to\infty$.  Indeed, we note that saturation of the black disk limit can be modelled by the following
rational representation 
\begin{equation}\label{bd}
f(s,b)=\tilde U(s,b)/[1+2\tilde U(s,b)],
\end{equation}
where the function $\tilde U(s,b)$ has similar to the function $ U(s,b)$ dependencies on  $s$ and $b$. Eq. (\ref{darb}) can be used then for the restoration of the real part in the case of the black disk limit saturation.  

And, finally, for the real and imaginary parts of the elastic scattering amplitude $F(s,t)$ we have in the case  of the unitarity limit saturation at $s\to\infty$:
\begin{equation}\label{hel}
\frac{\mbox{Im} F(s,t)}{s}\sim \frac{r(s)J_1(r(s)\sqrt{-t})}{\sqrt{-t}}
\end{equation}
and
\begin{equation}\label{hinel}
\frac{\mbox{Re} F(s,t)}{s} \sim {r(s)J_0(r(s)\sqrt{-t})}.
\end{equation}
The latter relation results from the asymptotical peripheral dependence of $h_{inel}(s,b)$ with sharp maximum at $b=r(s)$.
Thus, at asymptotic energies, the ratio $\rho$ would depend on the both variables $s$ and $-t$ and should not  be a constant over $-t$. This takes place in the region of the small values of $-t$ also and conforms to the conclusion on existence of the nontrivial $-t$ dependence of the function $\rho$ in the CNI region \cite{martin,kund}. The reason can be traced to the different forms of the impact parameter dependence corresponding to the real and imaginary parts of the elastic scattering amplitude.  

The aim of the present note is to consider the case of unitarity saturation at $s\to\infty$. 
At small values of $-t$ the Bessel functions $J_0$ and $J_1$ can be approximated by the linear over $t$ exponential functions with {\it different slopes}.  This means that $d\sigma/dt$ at small-$t$  and $s\to \infty$  can, in principle, be approximated by the two exponential functions. The function $J_0$ is sharply peaked, it can be correlated with the inelastic processes' contribution.
The zeros of the functions  $J_0$ and $J_1$ would lead to the dips in $d\sigma/dt$. It should be noted that those simple forms of the real and imaginary part contributions are valid at the asymptotical energies only. But, this fact serves as a motivation for an extension of the simple exponential dependence by adding another linear exponent instead of changing linear exponent into a nonlinear one \cite{totem}. Another motivation for such parameterization  can be based on the account of spins of colliding protons \cite{krisch}. Since a long time fits to $d\sigma/dt$ with several exponential functions are known to be successful  at lower energies. 

\section{Additive fit of $d\sigma / dt$  in the low$-t$ region}
Thus, despite that  the  LHC energies are not the asymptotic ones,  the above asymptotic results to the LHC energies might happen to be at least in a  qualitative correspondence with the recent TOTEM result on the $7\sigma$ deviation of  the $d\sigma/dt$ at small-$t$ from a simple linear exponential dependence over $-t$ \cite{totem}. 

To be a more quantitative, we have performed a simple fit to the TOTEM data using sum of two linear exponential functions. The aim of this fit is  to illustrate existence of a working alternative  to the parameterization with a quadratic exponential function\footnote{Such parameterization of the differential cross-section implies the equal slopes of the real and imaginary parts of the scattering amplitude at $t=0$.  This form should be considered as a deficient one owing to this equality of slopes it implies. Such conclusion results from the derivative dispersion relations for the amplitude slopes \cite{ferreira}, the new experimental  results on the amplitude phase \cite{totcni} and the above discussion.}  $Ae^{Bt-Ct^2}$used in \cite{totem}. The following function: 
\begin{equation}\label{fit}
\frac{d\sigma}{dt}=A_1e^{B_1t}+A_2e^{B_2t}
\end{equation}
has been used.
The resemblance with the results described above is a qualitative one and a similar form might  also be motivated by an account for  the spin amplitudes\footnote{If the spin dependence in elastic scattering would survive at such high energies, one can use then a rather complicated  fit, which includes ten linear exponential functions according to the presence of the five independent helicity amplitudes in $pp$-scattering description.}.
 The description with two linear exponents appears to be successful with the following values of the fitting parameters: $A_1=524.075$ mb/GeV$^2$, $A_2=10.036$ mb/GeV$^2$ and
$B_1=19.897$ GeV$^{-2}$ and $B_2=9.950$ GeV$^{-2}$. The value of $\chi^2/ndf$ is 0.46. The table data from \cite{totem} have been used and their description is shown in Fig. 1. 
\begin{figure}[hbt]
\begin{center}
\resizebox{9cm}{!}{\includegraphics*{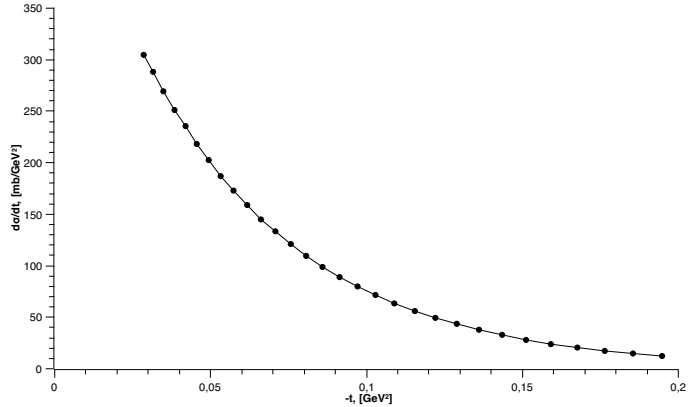}}
\end{center}
\caption[ch1]{\small Description of the differential cross-section of elastic $pp$-scattering in the limited region of low values of $-t$ at $\sqrt{s}=8$ TeV, the data have been taken from \cite{totem}.}
\end{figure}
\section*{Conclusion}
The restoration scenario of the real part of the elastic scattering amplitude has been considered in the case of the unitarity saturation at $s\to\infty$. 
The unitarity saturation is indicated by the recent experimental data at the LHC energy range.

The real part of the amplitude appears to have a  different $t$-dependence compared to its imaginary part in this particular case. The difference in the dependencies of the real and imaginary parts of the scattering amplitude means presence of a nontrivial $t$--dependence in the amplitude phase and can be correlated with  the recent TOTEM results on the measurements of $d\sigma /dt$ in the low-t region at $\sqrt{s}=8$ TeV. It evidently also results in  the $t$-dependent ratio of the real to the imaginary parts of the elastic scattering amplitude. 

Numerical fit with sum of two linear exponents having different slopes has been performed to illustrate consistency of the experimental results with additivity  of the contributions into the differential cross-section $d\sigma/dt$ as it was discussed above. 
\section*{Acknowledgements}
We are grateful to V.V. Anisovich for the stimulating correspondence.


\begin{thebibliography}{99}
\bibitem{totem}
G. Antchev et al. (The TOTEM Collaboration),  Nucl. Phys. B. {\bf 899},  527 (2015). 
\bibitem{martin}
A. Martin,  Phys. Lett. B {\bf 404}, 137 (1997). 
\bibitem{khuri}
N.N. Khuri, Proc. of Vth Blois Workshop --- International Conference on Elastic
 and Diffractive Scattering, Providence, RI, 8-12 Jun 1993, Editors H.M. Fried, K. Kang and
 C-I Tan, World Scientific (Singapore), 1994 p. 42.
\bibitem{bour}
C. Bourerely, N.N. Khuri, A. Martin, J. Soffer, T.T. Wu, Contribution to the proceedings of the XIth International Conference on Elastic and Diffractive Scattering, Ch\^ateau de Blois, May 15-20, 2005, presented by T.T. Wu.
\bibitem{totcni}
G. Antchev et al. (The TOTEM Collaboration), arXiv: 1610.00603.
\bibitem{trosh}
S.M. Troshin, Phys. Lett. B {\bf 682},  40 (2009). 
\bibitem{anis2}
V.V.  Anisovich, V.A. Nikonov, J. Nyiri, Int. J. Mod. Phys. A {\bf 30},   1550188 (2015).
\bibitem{white}
I.M. Dremin, S.N. White, arXiv:1604.03469.
\bibitem{grib}
V.N. Gribov,  A.A. Migdal, Sov. J. Nucl. Phys. {\bf 8},  583 (1969).
\bibitem{bronz}
J.B. Bronzan, ANL/HEP-7327, 33 (1973) .
\bibitem{bronz1}
J.B. Bronzan, G.L. Kane, U.P. Sukhatme, Phys. Lett. B {\bf 49}, 272 (1974).
\bibitem{ph1}
S.M. Troshin, N.E. Tyurin, Phys. Lett. B {\bf 208},  517  (1988).
\bibitem{phl}
S.M. Troshin, N.E. Tyurin, Phys. Lett. B {\bf 316},   175 (1993).
\bibitem{degr}
P. Desgrolard, L.L. Jenkovszky, B.V. Struminsky, 
Phys. Atom. Nucl. {\bf 63},  891 (2000).
\bibitem{intja}
S.M. Troshin, N.E. Tyurin, Int. J. Mod. Phys. A {\bf 22},  4437 (2007).
\bibitem{anis}
V.V.  Anisovich, V.A. Nikonov, J. Nyiri,
Phys. Rev. D {\bf 90}, 074005  (2014).
\bibitem{fagundes}
D.A. Fagundes, M.J. Menon, P.V.R.G. Silva, Nucl. Phys. A 946, 194 (2016).
\bibitem{alkin}
A. Alkin, E. Martynov, O. Kovalenko, and S.M. Troshin, { Phys. Rev. D} {\bf 89}, 091501(R) (2014) .
\bibitem{8tev}
G. Antchev et al. (The TOTEM Collaboration)
Phys. Rev. Lett. {\bf 111},  012001 (2013).
\bibitem{refl}
S.M. Troshin, N.E. Tyurin, Mod. Phys. Lett. {\bf 31}, 1650025 (2016).
\bibitem{drem}
I.M. Dremin, arXiv:1605.08216.
\bibitem{granad}
J.L. Albacete, A. Soto-Ontoso,  arXiv:1605.09176.
\bibitem{kinosh}
N.N. Khuri, T. Kinoshita, Phys. Rev. {\bf 137},  B720 (1965).
\bibitem{heitler}
W. Heitler, Proc. Camb. Phil. Soc. {\bf 37},  291 (1941).
\bibitem{mpla16}
S.M. Troshin, N.E. Tyurin, Mod. Phys. Lett. A {\bf 31}, 1650079  (2016).
\bibitem{usmpla}
S.M. Troshin, N.E. Tyurin, Mod. Phys. Lett. A {\bf 24}, 1103 (2009).
\bibitem{kund}
V. Kundrat, M. Lokaji\v cek, Phys. Rev. D {\bf 31},  1045 (1985).
\bibitem{krisch}
A.D. Krisch, Phys. Rev. Lett. {\bf 19},  1149 (1967).
\bibitem{ferreira}
E. Ferreira, Int. J. Mod. Phys. E {\bf 16}, 2893 (2007).
\end{thebibliography}
\end{document}